\documentclass[a4paper,12pt]{article}
\usepackage[utf8]{inputenc}
\usepackage{amsmath}
\usepackage{amssymb}
\usepackage{epsf}
\usepackage{graphicx}
\usepackage{color}
\usepackage{indentfirst}
\usepackage{cite}
\usepackage{mathrsfs}

\begin{document}

    \begin{center}
        {\textbf{Time reversal invariance violation for high energy charged baryons in bent crystals}}
    \end{center}

    \bigskip

    \begin{center}
        \textbf{V.G. Baryshevsky}
    \end{center}

    \begin{center}
        {Research Institute for Nuclear Problems, Belarusian State
            University, \\ 11 Bobruiskaya str., 220030, Minsk, Belarus}
    \end{center}

\begin{abstract}
Spin precession of channeled particles in bent crystals at the LHC
gives unique possibility for measurements as electric and magnetic
moments of charm, beauty and strange charged baryons so and
constants determining  CP ($T_{odd}, P_{odd}$) violation
interactions and $P_{odd}, T_{even}$ interactions of baryons with
electrons and nucleus (nucleons).
%
%
For a particle moving in a bent crystal a new effect caused by
nonelastic processes arises: in addition to the spin precession
around the direction of the effective magnetic field (bend axis),
the direction of electric field and the direction of the particle
momentum, the spin rotation to the mentioned directions also
appears.
\end{abstract}

\noindent \textbf{\small{Crystal, bent crystal, charm baryon, electric dipole moment, spin rotation, parity violation, magnetic
moment, CP violation.}}

\section{Introduction}
The spin rotation phenomenon of channelled particles, moving in a
bent crystal, which was theoretically predicted in \cite{b1} and
observed in \cite{b2,b3,b4}, gives us the opportunity to measure
anomalous magnetic moment of high energy short-lived particles.
The appearance of beams with energies up to $7~Tev$ on LHC and
further growth of particles' energy and beams' luminosity on FCC
do essentially improve the possibility of using the phenomenon of
spin rotation of the high energy particles in bent crystals and
spin depolarization of such particles for measuring anomalous
magnetic moments of positively charged, as well as neutral and
negatively charged  short-lived hyperons, and $\tau$ -lepton
\cite{b5,b6,b7,b8}. The detailed analysis of conditions of the
experiment on measuring magnetic dipole moment (MDM) of charm
baryon $\Lambda^{+}_{c}$ on LHC, which has confirmed the
possibility of measuring MDM of such baryon on LHC, was
accomplished recently in \cite{b9}.
Strong electric field affects the channelled particle in a bent
crystal. As a consequence, the spin rotation phenomenon of the
channelled particle allows to obtain information about the
possible value of the electric dipole moment of short-lived
baryons, which elementary particles can obtain as a result of the
violation of the T-invariance \cite{b10, bn11}.

It should be noted, that besides electromagnetic interaction the
channelled particle moving in a crystal experiences weak
interaction with electrons and nuclei as well as strong
interaction with nuclei. Mentioned interactions lead to the fact,
that in the analysis of the particle's spin rotation, caused by
electric dipole moment interaction with electric field, both
$P_{odd}, T_{even}$ and $P_{odd}, T_{odd}$ non-invariant spin
rotation, resulting from weak interaction should be considered
\cite{b11,b12}.
 %

As obtained here, spin precession of channelled particles in bent
crystals at the LHC gives unique possibility for measurements as
electric and magnetic moments of charm, beauty and strange charged
baryons so and constants determining  CP ($T_{odd}, P_{odd}$)
violation interactions and $P_{odd}, T_{even}$ interactions of
baryons with electrons and nucleus (nucleons). A new effect arises
for a particle moving in a bent crystal due to nonelastic
processes, namely: along with the spin precession around three
directions (the direction of the effective magnetic field (bend
axis), the direction of electric field and the direction of the
particle momentum) the spin rotation to the mentioned directions
also appears.

 %
\section{Spin rotation and scattering particles in crystal}

The spin rotation phenomenon for high-energy particles, moving in
a bent crystal, as a result of quasi-classical motion  of
particles  channelled in crystals, can be described by equations
similar to those for motion of particles' spin in the storage ring
with the inner target \cite{b11,b12}.
The theory, which describes motion of the particle spin in
electromagnetic fields in a storage ring, has been developed in
many papers \cite{b13,b14,b15,b16,b17,b18,b19}.

According to \cite{b13,b14,b15,b16,b17,b18,b19},
the basic equation, which describes particle spin motion in an
electromagnetic field, is the Thomas-Bargmann--Michel--Telegdi
(T-BMT) equation. Refinement  of the T-BMT equation, allowing us
to consider the possible presence of the particle EDM, was made in
\cite{b20,b21}.

Now let us consider a particle with  spin $S$ which  moves in the
electromagnetic field.
The term ''particle spin'' here means the expected value of the
quantum mechanical spin operator $\hat{\vec{S}}$ (hereinafter the symbol marked with "hat"
means a quantum mechanical operator).
Further, movement of the high-energy particles in non-magnetic crystal will be considered. In this case  magnetic field $ \vec{B}=0 $ and Lorentz-factor $ \gamma\gg 1 $.

Spin motion is described by the Thomas--Bargmann--Michel--Telegdi
equation (T-BMT) in this case as follows:
\begin{equation}
\frac{d \vec{S}}{d t}=[\vec{S}\times\vec{\Omega}]\,,
\label{eq1}
\end{equation}
\begin{equation}
\vec{\Omega}=-\frac{e}{mc} a \left[\vec{\beta}\times\vec{E} \right],
\label{eq2}
\end{equation}
where $\vec{S}$ is the spin vector, $t$ is the time in the
laboratory frame,$m$ is the mass of the particle, $e$ is its charge,  $\gamma$ is
the Lorentz-factor,$\vec{\beta}=\vec{v}/c$, where $\vec{v}$ denotes the particle velocity, $\vec{E}$is the electric field at the point of particle location in the laboratory frame, $a=(g-2)/2$ and $g$ is the gyromagnetic ratio
(by definition, the particle magnetic moment $\mu=(eg\hbar/2mc)S$, where $S$ is the particle spin).
The T-BMT equation describes the spin motion in the rest frame of
the particle, wherein the spin is described by the three component
vector $\vec{S}$.
In practice the T-BMT equation works well for the description of
spin precession in the external electric and magnetic fields
encountered in typical present--day accelerators.
Study of the T-BMT equation allows us to determine the major
peculiarities of spin motion in an external electromagnetic field.
However, it should be taken into account that particles in an
accelerator or bent crystal have an energy spread and move along different orbits.
This necessitates one to average the spin--dependent parameters of
the particle over the phase space of the particle beam.
This is why one must always bear in mind the distinction between
the beam polarization $ \vec{\xi} $ and the spin vector $\vec{S}$.
A complete description of particle spin motion can be made
applying spin density matrices equation (in more details see
\cite{b12,b22}).

If a particle possesses an intrinsic electric dipole moment, then
the additional term, describing spin rotation induced by the EDM,
should be added to (\ref{eq1}):
\begin{equation}
\frac{ d \vec{S}_{\mathrm{EDM}}}{  d  t}=\frac{ d }{S\hbar}
\left[\vec{S}\times\left\{\vec{E}-\frac{\gamma}{\gamma+1}\vec{\beta}(\vec{\beta}\vec{E})\right\}\right]\,,
\label{eq3}
\end{equation}
where $d$ is the electric dipole moment of the particle.

As a result, the motion of particle spin due to the magnetic
and electric dipole moments can be described by the following
equation:
\begin{eqnarray}
\frac{ d \vec{S}}{d t} & =& -\frac{e}{mc} a \left[\vec{S} \times\left[\vec{\beta}\times\vec{E} \right]\right] + \nonumber \\
&+&\frac{d}{\hbar S}\left[\vec{S}\times\left\{\vec{E}-
\frac{\gamma}{\gamma+1}\vec{\beta}(\vec{\beta}\vec{E})\right\}\right]\,.
\label{eq4.1}
\end{eqnarray}
Recall now, that electric field in a crystal is formed by atoms. Scattering on atoms leads to the fact, that the high-energy particle moving in a crystal experience interaction from electric and magnetic fields. However it is not only the electromagnetic interaction that influence on the scattering. Particles also participate in strong and weak interactions with electrons and nuclei.The interactions, mentioned above, depend on the spin of colliding particles and therefore have effect on evolution of the spin of the particle moving in matter \cite{b12}.

It would be recalled that the particle refractive index in matter formed by different scatterers has the form:
\begin{equation}
n=1+\frac{2\pi N }{k^{2}}f\left( 0\right)\,,
\label{eq5.1}
\end{equation}
where $N$ is the number of scatterers per $cm^3$ and $k$ is the wave number of the particle incident on the target,
$f(0)\equiv f_{aa} (\vec{k}' -\vec{k}=0)$ is the coherent elastic zero angle scattering amplitude.
%
In this scattering,  momentum of the scattered particle $\vec{p} '=\hbar\vec{k}'$ ( where $\vec{k}'$ is a wave vector) equals to initial momentum $\vec{p}=\hbar\vec{k}$. Atom (nucleus) that was in quantum state before interaction with the incident particle characterized by stationary wave function $\Phi_{a}$ will stay in the same quantum state after interaction with the incident particle.
%
%
%
If the energy of interaction between particle and a scatterer depends on spin of the particle, then scattering amplitude $\hat{f}(\vec{k}' -\vec{k})$ can also depend on spin. As a consequence refractive index $\hat{n}$ (symbol $\hat{}$ means that mentioned magnitude is an operator in spin space of a particle) depends on spin as well \cite{b12}.

If the matter is formed by different scatterers, then
\begin{equation}
n=1+\frac{2 \pi}{k^2}\sum_{j}N_j f_j (0),
\label{eq4}
\end{equation}
where $N_j$ is the number of j-type scatterers per $cm^3$,  $f_j (0)$ is the amplitude of the particle coherent elastic zero-angle scattering by j-type scatterer.

Let us consider a relativistic particle refraction on the
vacuum-medium boundary (see \cite{b12}). The wave number of the
particle in the vacuum is denoted $k$. The wave number of the
particle in the medium is $\vec{k}'=kn$. As is evident the
particle momentum in the vacuum $\vec{p}=\hbar\vec{k}$ is not
equal to the particle momentum in the medium. Therefore, the
particle energy in the vacuum $E=\sqrt{\hbar^2 k^2 c^2 +m^2 c^4}$
is not equal to the particle energy in the medium
$E_{med}=\sqrt{\hbar^2 k^2 n^2 c^2 +m^2 c^4}$.

The energy conservation law immediately requires the particle in
the medium to have the effective potential energy $U_{eff}$. This
energy can be easily found from relation:

\begin{equation}
E=E_{med}+U_{eff},
\label{eq5}
\end{equation}
i.e.
\begin{equation}
U_{eff}=E-E_{med}= -\frac{2 \pi \hbar^2}{m \gamma} N f(E,0)= (2
\pi)^3 NT_{aa} (\vec{k}' -\vec{k}=0),
\label{eq5a}
\end{equation}
\begin{equation}
f(E,0)=-(2 \pi)^2 \dfrac{E}{c^2 \hbar^2} T_{aa} (\vec{k}'
-\vec{k}=0) = -(2 \pi)^2 \frac{m \gamma}{\hbar^2}T_{aa} (\vec{k}'
-\vec{k}=0),
\label{eq5b}
\end{equation}
%
%
where $T_{aa} (\vec{k}' -\vec{k}=0)$ is the matrix element of the T-operator describing elastic coherent zero-angle scattering .

Let us remind that T-operator  is associated with scattering matrix $S$ \cite{bn13, bn15}:
 \begin{equation}
S_{ba}=\delta_{ba} -2 \pi i \delta (E_{b}-E_{a}) T_{ba},
 \label{eq6}
 \end{equation}
%
where $E_a$ is the energy of scattered particles before the collision, $E_b$ - after the collision, matrix element $T_{ba}$ corresponds to states $ a $ and $ b $ that refer to the same energy.

For the matter formed by different scatterers effective potential energy can be written as:
 \begin{equation}
U_{eff}=- \dfrac{2 \pi \hbar^2}{m \gamma} \sum_{j}N_j f_j (E,0).
\label{eq7}
\end{equation}
Due to periodic arrangement of atoms in a crystal the effective potential energy is a periodic function of coordinates of a particle moving in a crystal \cite{b12}:

\begin{equation}
U(\vec{r})=\sum_{\vec{\tau}}U(\vec{\tau}) e^{i \vec{\tau} \vec{r}},
\label{eq8}
\end{equation}
where $\vec{\tau}$ is the reciprocal lattice vector of the crystal;

\begin{equation}
U(\vec{\tau})=\dfrac{1}{V}\sum_{j}U_j (\vec{\tau}) e^{i \vec{\tau} \vec{r_j}},
\label{eq9}
\end{equation}
here $V$ is the volume of the  crystal elementary cell,
$\vec{r_j}$ is the coordinate of the atom (nucleus) of type $ j $
in the crystal elementary cell.

\begin{equation}
U_{j}(\vec{\tau})=- \frac{2 \pi \hbar^2}{m \gamma} F_j (\vec{\tau}),
\label{eq10}
\end{equation}
%

According to (\ref{eq10}) effective potential energy
$U(\vec{\tau})$ is determined by amplitude $F_j
(\vec{\tau})=F_{jaa}(\vec{k}' -\vec{k}=\vec{\tau}).$ In contrast
to the case of chaotic matter where effective potential energy is
determined by the amplitude of elastic coherent scattering
$f(\vec{k}' -\vec{k})$, here it is defined by the amplitude $F
(\vec{\tau})$ (see Annex and \cite{b12}), which can be written as:
\begin{equation}
F_{j}(\vec{k}' -\vec{k})= f_{j}(\vec{k}' -\vec{k})-i\frac{k}{4 \pi}\int f_{j}^*(\vec{k}'' -\vec{k}')f_{j}(\vec{k}'' -\vec{k}) d\Omega_{ k''}.
\label{eq12}
\end{equation}
%
%
%
%
where $d\Omega_{ k''}$ means integration over all of the  vector $\vec{k}''$ directions,$|\vec{k}'|=|\vec{k}|=|\vec{k}''|$.

The occurrence of the amplitude $F(\vec{k}' -\vec{k})$ instead of
the amplitude of elastic coherent scattering  $f(\vec{k}'
-\vec{k})$ in (\ref{eq6}) is specified by the fact, that unlike of
an amorphous matter, the wave elastically scattered in a crystal,
due to rescattering by periodically located centers is involved in
formation of a coherent wave propagating through the crystal.

As follows from (\ref{eq12}):

\begin{equation}
F_{j}(0)= f_{j}(0)-i\frac{k}{4 \pi}\int f_{j}^*(\vec{k}'' -\vec{k}')f_{j}(\vec{k}'' -\vec{k}) d\Omega_{ k''}.
\label{eq13}
\end{equation}
%
The integral in (\ref{eq13}) is identical with the total cross-section of the elastic coherent scattering by nucleus (atom). According to optical theorem:

\begin{equation}
Im f_{j}(0)= \frac{k}{4 \pi} \sigma_{tot}=\frac{k}{4 \pi} \sigma_{elast} + \frac{k}{4 \pi} \sigma_{nonelast}.
\label{eq14}
\end{equation}
%
As we can see, unlike the matter where scatterers are spread
chaotically in crystal, for the amplitude $ F_j (0) $
 following expression is true
\begin{equation}
F_{j}(0)=\tilde{f}_{j}(0),
\tilde{f}_{j}(0)= f_{j}(0)-\frac{k}{4\pi} \sigma_{elast}.
\label{eq15}
\end{equation}
%
In other words, cross-section of elastic coherent scattering in
crystal does not contribute to the imaginary part of the amplitude
$ F_j(0)$. Imaginary part is determined  by the cross-section of
nonelastic processes (reaction cross-section) only:
\begin{equation}
F_{j}(0)=Re F_{j}(0)+ i Im F_{j}(0) = Re F_{j}(0)+i\frac{k}{4 \pi} \sigma_{nonelast}.
\label{eq16}
\end{equation}

The situation is also similar for nonzero-angle scattering. It becomes clear when we use the equality that is correct for elastic scattering \cite{bn14}:
\begin{equation}
Im f_{elast} (\vec{k}' -\vec{k})= \frac{k}{4 \pi} \int f_{elast}^*(\vec{k}'' -\vec{k}')f_{elast}(\vec{k}'' -\vec{k}) d\Omega_{ k''}.
\label{eq17}
\end{equation}
%
%
As a result, according to (\ref{eq12}), we should subtract the elastic scattering contribution from the imaginary part of $ f(\vec{k}' -\vec{k})$ .
This can be evidently shown if the interaction with scatterer can
be considered in terms of the perturbation theory. In this case at
the first Born approximation scattering amplitude
$f^{(1)}(\vec{k}'-\vec{k})$ doesn't have an imaginary part:

\begin{equation}
Im f^{(1)}_{aa} (\vec{k}' -\vec{k})=0.
\label{eq18}
\end{equation}
%
An imaginary part appears at the second Born approximation.  Lets
remind that T-operator, determining the scattering amplitude (see
(\ref{eq5b})), satisfies the following equation \cite{bn13, bn15}:

\begin{equation}
T=V+V\frac{1}{E-H_{0}+i\eta}T.
\label{eq19}
\end{equation}
%
%
where $ V $ is the interaction energy, $ H_{0} $ is the Hamilton operator of colliding systems located at great distance from each other.

As a result for the amplitude of the elastic coherent scattering $ f_{aa} $ with the accuracy up to second order terms over the interaction energy, we have:

\begin{equation}
f_{aa} (\vec{k}' -\vec{k})= -(2 \pi)^2 \frac{m \gamma}{\hbar^2} (<\Phi_{\vec{k}'a}| V |\Phi_{ka}> + <\Phi_{\vec{k}'a}| V \frac{1}{E_{a} (\vec{k})-H_{0}+i \eta} V |\Phi_{ka}>),
\label{eq20}
\end{equation}
%
 where $ \Phi_{ka} $ is an eigenfunction of Hamilton operator $ H_{0}$,
\begin{equation}
\Phi_{\vec{k}a}=\frac{1}{(2 \pi)^{3/2}} e^{i \vec{k} \vec{r}}\Phi_{a},
\label{eq200}
\end{equation}
%
%
$\Phi_a $ is a wave function of scatterer
stationary states, $H_{0}\Phi_{\vec{k}a}=E_{a}(\vec{k})\Phi_{ka}$.
Using the completeness of the function $\Phi_{\vec{k}a} $ and
replacing "1" in (\ref{eq20}) by
$\sum_{k''b}|\Phi_{k''b}><\Phi_{k''b}|=1$, in (\ref{eq20}) we obtain
the sum over the intermediate states $b$,  which includes states
with $ b=a $. This term contains the following expression:

\begin{equation}
\frac{1}{E_{a}(\vec{k})-E_{a}(\vec{k}'')+i\eta}=
P \frac{1}{E_{a}(\vec{k})-E_{a}(\vec{k}'')}
- i \pi \delta (E_{a}(\vec{k})-E_{a}(\vec{k}'')).
\label{eq21}
\end{equation}
%
%
The $"P"$ symbol in the real part of (\ref{eq21}) means, that in (\ref{eq20}) integrals containing the  $"P"$ symbol are principal-value integrals.
%

This real part contribution to the first Born approximation is
negligibly small therefore it will not  be considered further. The
imaginary unit in second term of (\ref{eq21}), which is
proportional to the $ \delta $ function, leads to occurrence of an
imaginary part in amplitude $ f (\vec{k}' -\vec{k}) $. After
substitution of the expression with $ \delta $ function into
(\ref{eq20}) it becomes obvious that the term in which sum $ b=a $
is equivalent to the term subtracted from the amplitude  $ f_{aa}
(\vec{k}' -\vec{k}) $ in (\ref{eq12}). As a result only
contributions caused by nonelastic processes and reactions with
$b\neq a$  make contributions to the imaginary part of the
amplitude in crystal. Further in expressions for $ F $  amplitude
$ f_{aa}$  without the contribution of the elastic coherent
scattering in the imaginary part will be considered. The second
term in (\ref{eq12}) also will not be written explicitly.

\section{Effective potential energy of a spin-particle moving close to crystal planes (axes)}

Suppose a high energy particle moves in a crystal at a small angle
to the crystallographic planes (axes) close to the Lindhard angle.
This motion determined by the plane (axis) potential $ \vec{U}(x)
(\vec{U}(\vec{\rho}))$, which can be determined from
$\vec{U}(\vec{r})$ by averaging over distribution of atoms
(nuclei) in a crystal plane (axis).
%
Similar result is obtained when all the terms with $ \tau_y \neq 0, \tau_z \neq 0  $ for the case of planes  or $ \tau_z \neq 0 $ for the case of axes are removed from the sum (\ref{eq8}).

As a consequence for the potential of periodically placed axes we can write:

\begin{equation}
U(\vec{\rho})=\sum_{\vec{\tau}_{\perp}}U(\vec{\tau}_{\perp},\vec{\tau}_{z}=0) e^{i\vec{\tau}_{\perp}\vec{\rho}},
\label{eq172}
\end{equation}
$ z $-axis of the coordinate system is directed along the crystallographic axis. For the potential of a periodically placed planes we have:

\begin{equation}
U(x)=\sum_{\vec{\tau}_{x}}U(\tau_{x},\tau_{y}=0, \tau_{z}=0)
e^{i\tau_{x}x},
\label{eq182}
\end{equation}
%
$y,z $-planes of the coordinate system are parallel to the chosen
crystallographic planes family. Lets remind that according to
(\ref{eq9}-\ref{eq10}) the magnitude $U(\vec{\tau})$ is expressed
in terms of the amplitude $ F(\vec{\tau}) $.

Since the amplitude $ \hat{F}(\vec{k}'-\vec{k}) $ depends on spin,
the effective potential energy $\hat{U}$ depends on a spin as well
\cite{b12}. The magnitude $\hat{U}$ is the operator in spin space
of a particle incident on a crystal.

Elastic coherent scattering of a particle by an atom is caused by
electromagnetic interaction of the particle with the atom
electrons and nucleus as well as weak and strong nuclear
interaction  with electrons and nucleus.The general expression for
an amplitude of elastic scattering of spin $\frac{1}{2}$ particles
by a spinless or unpolarized nuclei can be written as:

\begin{eqnarray}
\label{eq182a}
\hat{F}(\vec{q})=A_{coul}(\vec{q})+A_{s}(\vec{q})+( B_{magn}(\vec{q}) + B_{S}(\vec{q}))\vec{\sigma}
[\vec{n}\times\vec{q}]+ \\
+( B_{we}(\vec{q})  + B_{w nuc}(\vec{q}))\vec{\sigma}\vec{N}_{w} + \nonumber \\
+ (B_{EDM}(\vec{q}) + B_{Te}(\vec{q}) +
B_{Tnuc}(\vec{q}))\vec{\sigma}\vec{q}, \nonumber
\end{eqnarray}
%
where $\vec{q}=\vec{k}'-\vec{k}, \vec{n}= \frac{\vec{k}}{k} , A_{coul} (\vec{q})$ is the spin-independent part of the amplitude of elastic coulomb scattering of a particle by an atom (according to (\ref{eq172},\ref{eq182})) it leads to Coulomb potential energy of crystal planes and axes);  $A_{s}(\vec{q})$ is the spin-independent part of a scattering amplitude, which is caused by strong interaction (similar contribution caused by weak interaction is negligibly small therefore it is omitted).

%
The spin-dependent amplitude, which is proportional to $B_{magn}(\vec{q})$, is determined by electromagnetic spin-orbit interaction.This amplitude is responsible for the effect of particle spin rotation in the electric field of a bent crystal, that is proportional to $(g-2)$. The term proportional to $B_{s} (\vec{q})$ is responsible for the contribution of  the spin-orbit strong interaction to a scattering process of a baryon by nucleus.This term leads to spin rotation caused by strong interaction.
%

The term proportional to the parity odd pseudo scalar $ \vec{\sigma} \vec{N}_{w}$ (unit vector $ \vec{N_w}=\frac{\vec{k}' + \vec{k}}{|\vec{k}'+\vec{k}|}$) includes two contributions:
a) Contribution to the amplitude proportional to $B_{we}(q)$, that describes elastic scattering caused by the parity violating
weak interaction between baryon and electrons.
b)  Contribution to the amplitude proportional to $ B_{wnuc}(\vec{q})$,that describes elastic scattering caused by the parity violating weak interaction between baryon and nucleus.
%

The term proportional to the time (T) violation (CP non-invariant)
pseudo scalar $ \vec{\sigma}\vec{q}$  includes three
contributions: a) Contribution proportional to $ B_{EDM} (q) $,
that describes elastic scattering of baryon with EDM by the atom's
Coulomb field. This contribution leads to the term in
(\ref{eq4.1},\ref{eq172},\ref{eq182}) describing baryon spin
rotation in electric field of planes (axes) caused by EDM. b)
Contribution proportional to $ B_{Te} (q) $ describes possible
short-distance T-non-invariant interaction between baryon and
electrons. c) Contribution to the amplitude, which is proportional
to $ B_{Tnuc} (q) $, describes scattering caused by
T-non-invariant interaction between baryon and nucleons. This
contribution also leads to spin rotation of a baryon moving in
bent crystal.
%

Therefore it is also possible to examine T-non-invariant
baryon-nucleon interaction (and short-range baryon-electron
interaction) and  obtain restrictions on the value of such
interactions in experiments on measurements of charm and beauty
baryons EDM.


Lets express the amplitude $\hat{F}(q)$ as Fourier transformation of function $\hat{F} (\vec{r})$:
 \begin{equation}
 \hat{F}(\vec{q})=\int \hat{F}(\vec{r}') e^{-i \vec{q}\vec{r} '} d^3 r'.
 \label{eq22}
 \end{equation}
%
Considering mentioned above we can conduct summation of $\tau_x$ and $\vec{\tau_{\perp}}$ in (\ref{eq172},\ref{eq182}) using following expression:
 \begin{equation}
 \sum_{\tau_{x}} e^{i\tau x}=d_{x} \sum_{l} \delta(x-X_{l}),
 \label{eq23}
 \end{equation}
%
where $ d_x $ is the lattice period along axis $x$; $ X_l $ are coordinates of $ l $ plane.
\begin{equation}
\sum_{\tau_{x}, \tau_{y}} e^{i\tau_{\perp}\vec{\rho}}=d_{x} d_{y} \sum_{l} \delta(\vec{\rho}-\vec{\rho}_{l}),
\label{eq24}
\end{equation}
%
where $\vec{\rho_l}$ is a coordinate of an axis, located in point $\vec{\rho_l}$; $d_x, d_y$ are lattice periods along axes $x$ and $y$.

As a result we obtain following expression for the effective potential energy of interaction between an incident particle and a plane (axis) (the lattice is assumed to consist of atoms of one kind):
\begin{equation}
\hat{U}(x) = - \sum_{\tau_{x}}\frac{2 \pi \hbar^2 }{m \gamma V} \hat{F} (q_{x}=\tau_{x}, q_{y}=q_{z}=0) e^{i\tau_{x}x}=
-\frac{2 \pi \hbar^2 }{m \gamma V d_{y} d_{z}} \hat{F} (x, q_{y}=q_{z}=0),
\label{eq25}
\end{equation}

\begin{equation}
\hat{U}(\vec{\rho}) = -\frac{2 \pi \hbar^2 }{m \gamma V}
\sum_{\tau_{x}, \tau_{y}} \hat{F} (q_{x}=\tau_{x}, q_{y}=\tau_{y},
q_{z}=0)e^{i\tau_{\perp}\vec{\rho}}= -\frac{2 \pi \hbar^2 }{m
\gamma d_z} \hat{F} (\vec{\rho}, q_{z}=0),
\label{eq26}
\end{equation}
%
$d_z$ is the lattice period along the axis $z$.

Lets consider the expression for  effective potential energy in
detail. According to (\ref{eq172},\ref{eq25},\ref{eq26})
contributions to the effective potential energy are caused by
interactions of different types  including short-range and
long-range interactions.
%
In the presence of several types of interaction, for describing
their different contributions to the scattering amplitude as a
result of their mutual influence on scattering process, it is
convenient to separate scattering caused only by long-range
interactions and present amplitude in following form:
\begin{equation}
f(\vec{q})= f_{long} (\vec{q})+ f_{short long} (\vec{q}),
\label{eq27}
\end{equation}
%
where $f_{long} (\vec{q})$ is a scattering amplitude determined by long-range coulomb and magnetic interactions  (assuming that short-range interactions are absent), $f_{short long} (\vec{q})$ is a scattering amplitude determined by short-range interactions (calculating this amplitude waves scattered by long-range interactions were used as an incident waves). For general scattering theory in the presence of several interactions see, for example, \cite{bn13,bn15}.

The interactions mutual influence on a scattering amplitude can be
easily followed with the help of perturbation theory. Let
interaction energy $V$ a sum of several interactions: $V=\sum_i
V_i$. Then at the first Born approximation scattering amplitude is
a sum of scattering amplitudes caused by every interaction
separately: $f=\sum_i f_1 (V_i)$. But at the second  Born
approximation scattering amplitude is determined by squared
interaction $V$, more precisely by the following expression (see
\cite{bn13,bn14,bn15})
\begin{equation}
V \frac{1}{E-H_0 -i\eta}V = \sum_p V_p \frac{1}{E-H_0 -i\eta} \sum_l V_l ,
\label{eq28}
\end{equation}
%
As we can see, equality (\ref{eq28}) contains interference contributions to $f$ proportional to $V_p V_l$.

The coulomb amplitude, described by the first term in (\ref{eq182a}), leads to the usual expression for potential energy of interaction between a charged particle and  a plane (axis).

The second term $A_s (\vec{q}) $ is caused by short-range interaction. Amplitude $A_s (\vec{q})$ can be written as:

\begin{equation}
A_{s}(q)=A_{nuc} (q)\Phi_{osc} (\vec{q}),
\label{eq29}
\end{equation}
%
%
where $A_{nuc}(q)$ is the spin independent part of the amplitude of elastic scattering by the resting nucleus, $\Phi_{osc} (\vec{q})$ is the form-factor caused by nucleus oscillations in crystal.

Owing to the short-range kind of strong interactions amplitude
$A_{nuc} (\vec{q})$ is equal to zero-angle scattering amplitude
$A(0)$ in scattering angles range $\vartheta\leq
\frac{1}{kR_{osc}}\ll 1 $.

Form-factor $\Phi_{osc}(\vec{q})$ has the form \cite{bn14}:
\begin{equation}
\Phi_{osc}(\vec{q})=\sum_n \rho_n <\varphi_n (r)| e^{-i \vec{q}
\vec{r}} |\varphi_n (r)>= \int e^{-i \vec{q} \vec{r}} N_{nuc}
(\vec{r}) d^3 r,
\label{eq30}
\end{equation}
%
where $\varphi_n (r) $ is the wave function describing vibrational
state of nuclei in crystal, summation $\sum_{n}{\rho_n}$ means
statistical averaging with Gibbs distribution over vibrational
states of nucleus in crystal. Lets remind, that squared
form-factor $\Phi_{osc} (\vec{q}) $ is equal to Debye-Waller
factor, $ N_{nuc} (\vec{r}) $ is a probability density of
vibrating nuclei detection in point $\vec{r}$, $\int N_{nuc}
(\vec{r}) d^3r = 1 $.

As a result, according to (\ref{eq25}), this contribution to effective potential plane energy can be written as:
\begin{equation}
U_{nuc} (x)= -\frac{2 \pi \hbar^2}{m \gamma d_y d_z} N_{nuc} (x)
A_{nuc} (0),
\label{eq31}
\end{equation}
%
%
where $N_{nuc} (x)= \iint N_{nuc} (x, y', z') dy' dz'$ is the probability density of vibrating nuclei detection in point $x$ (in direction orthogonal to the chosen crystallographic plane).

Similarly, for the axis we have:
\begin{equation}
U (\vec{\rho})= -\frac{2 \pi \hbar^2}{m \gamma d_z} N_{nuc} (\vec{\rho}) A_{nuc} (0),
\label{eq32}
\end{equation}
where $N_{nuc} (\vec{\rho})= \int N_{nuc} (\vec{\rho}, z') dz'$.
\section{Effective potential energy determined by the anomalous magnetic moment}

According to (\ref{eq182a}) the scattering amplitude, which is determined by baryons magnetic moment, has the form:
\begin{equation}
\hat{F}_{magn} (q)= B_{magn} (q) \vec{\sigma} [\vec{n} \times \vec{q}].
\label{eq33}
\end{equation}
%

Defining the scattering amplitude one could at first consider only
magnetic scattering and its interference with coulomb scattering
(see (\ref{eq27})), and then add the term caused by interference
between magnetic and nuclear interactions.

For the first case perturbation theory can be used. In the first
order of perturbation theory interference between the magnetic
moment scattering by the coulomb field and the coulomb scattering
of baryon electric charge by the coulomb field is absent. The
amplitude $ \hat{F}^{(1)}$ can be written as:
\begin{equation}
\hat{F}_{magn}^{(1)} (\vec{q}) = i f_{coul} (\vec{q})\frac{\hbar}{m c} (\frac{g-2}{2})\frac{1}{2} \vec{\sigma} [\vec{n} \times \vec{q}] ,
\label{eq34}
\end{equation}
%
%
where  $f_{coul} (\vec{q})$ is the amplitude of coulomb scattering of a baryon by an atom in the first Born approximation; $ \vec{n}=\frac{\vec{k}}{k}$, $m$ is the baryon mass.

It should be noted that amplitude ${F}_{magn} (q)$  is a pure
imaginary quantity. After substitution of (\ref{eq34}) into
(\ref{eq25}) and summation over $\tau_x$ one obtains the
expression for effective interaction energy:
 \begin{equation}
 \hat{U}_{magn}(x) = -\frac{e \hbar}{2 m c} \frac{g-2}{2} \vec{\sigma} [\vec{E}_{plane}(x) \times \vec{n}] ,
 \label{eq35}
 \end{equation}
%
%
where $\vec{E}_{plane}(x)$ denotes the electric field, produced by
the crystallographic plane in point $x$. In axis case
$U_{magn}(\vec{\rho})$ can be obtained by replacement of $x$ by
$\vec{\rho}$ in (\ref{eq35}) and $\vec{E}_{plane} (x)$ by
$\vec{E}_{axis} (\vec{\rho})$.

Using (\ref{eq35}) and Heisenberg equations for spin operator, the motion equation for  polarization vector (\ref{eq1}, \ref{eq2}) for the case of  $ B=0 $ and $\gamma >>1 $ can be obtained.

The equation (\ref{eq35}) can be represented as:

\begin{equation}
\hat{U}_{magn} = -\frac{e \hbar}{2 m c} \frac{g-2}{2}E_{x plane}(x) \vec{\sigma} \vec{N},
\label{eq36}
\end{equation}
%
%
%
where $\vec{N}=[\vec{n}_x \times \vec{n}]$ is the unit vector , $\vec{n}_x \perp \vec{n}$, direction of the unit vector $\vec{n}$  is parallel to the crystallographic plane.

The expression (\ref{eq36}) for the effective potential energy is
purely real. However, the scattering amplitude $ \hat{F}(\vec{q})$
has both real and imaginary parts. Due to this fact, the effective
potential energy  $\hat{U}$ also has real and imaginary parts.

In the second order of perturbation theory the amplitude $ \hat{F}(\vec{q})$ is not purely imaginary, but also has a real part.
By means of (\ref{eq12}, \ref{eq20}-\ref{eq21}) the following expression for the contribution $\tilde{F}^{(2)} (q)$  to the amplitude $ \hat{F}(\vec{q})$ can be obtained:

\begin{eqnarray}
\label{eq37}
& & \tilde{F}^{(2)} (\vec{q}=\vec{\tau}) = i \frac{k}{4
    \pi\hbar^2 c^2} \left\{ <\Phi_a|\iint e^{-i \vec{\tau}
    \vec{r}_{\perp}}
\left[ \int \hat{V} (\vec{r}_{\perp},z)dz \right] ^2 d^2 r_{\perp} |\Phi_a>  - \right. \nonumber \\
& & \left. -  \iint e^{-i \vec{\tau} \vec{r}_{\perp}} \left[ \int
<\Phi_a| \hat{V}(\vec{r}_{\perp},z) | \Phi_a> dz \right]^2 d^2
r_{\perp} \right\} = \nonumber \\
& = & i \frac{k}{4 \pi \hbar^2 c^2} \iint e^{-i \vec{\tau}
    \vec{r}_{\perp}} \left\{ <\Phi_a| \left[ \int \hat{V}
(\vec{r}_{\perp},z)dz \right]^2 |\Phi_a> - \right.  \\
& & \left. -  \left[ \int <\Phi_a|
\hat{V} (\vec{r}_{\perp},z) |\Phi_a> dz \right]^2 \right\} d^2
r_{\perp}
\nonumber = \\
& = & i \frac{k}{4 \pi \hbar^2 c^2} \iint   e^{-i \vec{\tau}
    \vec{r}_{\perp}} \left\{
{\overline { \left[ \int \hat{V}(\vec{r}_{\perp},z)dz
        \right]^2} }
-  \Bigg[ \overline{\int \hat{V}(\vec{r}_{\perp},z)dz} \, \Bigg]^2
\right\} d^2 r_{\perp}, \nonumber
\end{eqnarray}
%
%
%
%
%
where $ \hat{V}(\vec{r}_{\perp},z)= \hat{V}_{coul}(\vec{r}_{\perp},z)+\hat{V}_{magn}(\vec{r}_{\perp},z)$,\\
$\hat{V}_{magn}(\vec{r}_{\perp},z)=- \mu_a  \vec{\sigma} [\vec{E}
(\vec{r}_{\perp}, z) \times \vec{n}]$, $z$ axis of the coordinate
system is directed along the unit vector $\vec{n}$, $\vec{n}$ is
the unit vector directed along the particle momentum before
scattering  $\hbar k $, $\mu_{a}$ is the anomalous magnetic moment
of the particle $\mu_{a}=\frac{e \hbar }{2 m c} (\frac{g-2}{2})$.

When deriving (\ref{eq37}), it was considered that  the particle energy is much greater than the electrons' binding energy in atoms and the atoms' binding energy in crystal.
As a result it is possible at first to examine scattering by
electrons and nuclei, which rest in points $r_i$,  and then to
average the result over electrons and nuclei positions with wave
functions $|\Phi_a>$ (impulse approximation, for example see
\cite{bn13}). The line in (\ref{eq37}) denotes such kind of
averaging.
The contribution caused by interference between magnetic and
nuclear scattering, and possible contributions determined by the
particle magnetic moment should complete the expression mentioned
above. In the case of positively charged particles, moving far
from the top of the potential barrier, the contribution caused by
interactions with nuclei is suppressed and will not be considered
in detail below.
After substitution of (\ref{eq37}) into (\ref{eq25}) and summation
over $\tau_x$ the following expression for the contribution to
the effective potential energy caused by the amplitude $
\tilde{F}_{magn} (\vec{\tau})$ can be obtained:
\begin{equation}
\hat{U}^{(2)}_{magn}(x)= -i \frac{1}{4 d_y d_z m c^2} (\frac{g-2}{2}) \frac{\partial}{\partial x} \overline{\delta V^2 (x)} \vec{\sigma}\vec{N},
\label{eq38}
\end{equation}
where $\vec{N}=[\vec{n}_{x} \times \vec{n}]$, $\vec{n}_{x} \perp \vec{n}, \vec{n}_{x} $ is the unit vector along axis $x$,
$ \overline {\delta V^2 (x)} = \int \left\{ \overline{ \left[\int V_{coul} (x,y,z) dz \right]^2 }   - \left[ \overline { \int V_{coul} (x,y,z) dz} \right]^2 \right\} dy $

Similarly for the case of axis it can be obtained:
\begin{equation}
\hat{U}^{(2)}_{magn}(\vec{\rho})= -i \frac{1}{d_z m c^2} (\frac{g-2}{2}) \vec{\sigma} [\nabla_{\rho} \overline{\delta V^2 (\vec{\rho})} \times \vec{n}].
\label{eq39}
\end{equation}
%
For the axisymmetric case:
\begin{equation}
\hat{U}^{(2)}_{magn}(\rho)= -i \frac{1}{4 d_z m c^2} (\frac{g-2}{2}) \frac{\partial}{\partial \rho}\overline{\delta V^2 (\rho)}
[\vec{n}_{\rho}  \times \vec{n}],
\label{eq40}
\end{equation}
$\vec{n}_{\rho}=\frac{\vec{\rho}}{\rho}$is the unit vector, $\vec{n}_{\rho} \perp \vec{n}$,
$\overline{\delta V^2 (\vec{\rho})} =  \overline{ \left[ \int V_{coul} (\vec{\rho},z)dz \right]^2}  - \left[\overline{\int V_{coul} (\vec{\rho},z)dz}\right]^2$.

Below the case of planar channeling will be considered. According to (\ref{eq36}, \ref{eq38}) the interaction between anomalous magnetic moment and plane can be written as:

\begin{eqnarray}
& \hat{U}_{magn}(x)&= \hat{U}_{magn}^{(1)}(x)+\hat{U}^{(2)}_{magn}(x)=  \\
&=&-\frac{e \hbar}{2 m c} \frac{g-2}{2} \vec{\sigma} [\vec{E}_{plane} \times \vec{n}]
-i \frac{1}{4 d_y d_z m c^2} (\frac{g-2}{2}) \frac{\partial}{\partial x}\overline{\delta V^2 (x)} \vec{\sigma}\vec{N}= \nonumber \\
&=&-(\alpha_m +i\delta_m)\vec{\sigma}\vec{N},\nonumber
\label{eq41}
\end{eqnarray}
where
\begin{eqnarray}
&&\alpha_m= \frac{e \hbar}{2 m c} \frac{g-2}{2} E_x, \nonumber \\
&&\delta_m= \frac{1}{4 d_y d_z m c^2} (\frac{g-2}{2}) \frac{\partial}{\partial x}\overline{\delta V^2 (x)}\nonumber
\end{eqnarray}
It will be shown herein, that the first term leads to spin rotation around $\vec{N}$, whereas the second term results in spin rotation in direction of $\vec{N}$.

\section{Effective potential energy $\hat{U}$ determined by spin-orbit interaction}

According to (\ref{eq182a}) the part of the scattering amplitude caused by the strong spin-orbit interaction has the form:
\begin{equation}
\hat{F}_{s sp-orb} (\vec{q}=\vec{\tau})=B_s (\vec{\tau}) \vec{\sigma}[\vec{n} \times \vec{\tau}].
\label{eq42}
\end{equation}
%
The coefficient $B_s (\vec{\tau}) $ can be expressed similar to (\ref{eq29}) as follows:
\begin{equation}
B_s (\vec{\tau})=B_{ nuc} (\vec{\tau})\Phi_{osc} (\vec{\tau}),
\label{eq43}
\end{equation}
%
where $B_{s nuc} (\vec{\tau})$ describes scattering by a resting nucleus, $\Phi_{osc}(\vec{\tau})$ is the form-factor determined by nucleus oscillations in crystal.

In the considered case, similar to the approach used when deriving
(\ref{eq31}), the short-range character of the nuclear forces and
small (as compared with the amplitude of nucleus oscillations)
nucleus radius  enables assumption $B_{s nuc} (\vec{\tau})\approx
B_{s nuc} (0)$. It is important that the coefficient $B_{s nuc}
(0)$ has both real and imaginary parts:
\begin{equation}
B_{s nuc} (0)=B_{s nuc}' + i B_{s nuc}''.
\label{eq44}
\end{equation}
This is similar to the case of amplitude, which describes scattering of the magnetic moment by the atom (nucleus).
To obtain the expression for the effective potential energy the
summation over  $\tau_x$ should be conducted in (\ref{eq25}). The
resulted expression is similar to that for $\hat{U}_{magn}$. For
example, for the plane case:
\begin{eqnarray}
&\hat{U}_{s sp-orb}&= -\frac{2 \pi \hbar ^2}{m \gamma d_y d_z} B_{s nuc} (0) (-i) \vec{\sigma}[\vec{n} \frac{\partial}{\partial \vec{x}} N_{nuc} (x)]= \nonumber \\
&=& -\frac{2 \pi \hbar ^2}{m \gamma d_y d_z} (B'' -iB')\frac{\partial N_{nuc}}{\partial x} \vec{\sigma} [\vec{n} \times \vec{n}_x]= \nonumber \\
&=& \frac{2 \pi \hbar ^2}{m \gamma d_y d_z} \frac{\partial N_{nuc}}{\partial x} B'' \vec{\sigma} \vec{N} - i \frac{2 \pi \hbar ^2}{m \gamma d_y d_z} B' \frac{\partial N_{nuc}}{\partial x} \vec{\sigma} \vec{N}.
\label{eq45}
\end{eqnarray}
Finally
\begin{equation}
\hat{U}_{s sp-orb}= -(\alpha_s +i\delta_s)\vec{\sigma} \vec{N},
\label{eq46}
\end{equation}
where
\begin{eqnarray}
&&\vec{N}= [\vec{n_x} \times \vec{n}],  \nonumber \\
&&\alpha_s=-\frac{2\pi \hbar^2}{m \gamma d_y d_z} \frac{\partial N_{nuc}}{\partial x} B'', \nonumber \\
&&\delta_s = \frac{2\pi \hbar^2}{m \gamma d_y d_z} B'
\frac{\partial N_{nuc}}{\partial x}. \nonumber \label{eq462}
\end{eqnarray}

As is shown below, similar to the magnetic contribution, the first
term, which is pure real, leads to rotation of spin around
$\vec{N}$, while the second pure imaginary term results in spin
rotation in the direction parallel or antiparallel to vector
$\vec{N}$.
Let us remind that the contribution determined by elastic
scattering, which is described by the second term in (\ref{eq13}),
should be subtracted from the expression for the amplitude.
However at high energies this contribution is negligibly small, in
comparison with nonelastic contributions to the amplitude, and
therefore can be omitted.

\section{Effective potential energy $\hat{U}$ determined by P-odd and T-even interactions}

The next group of terms, which are proportional to $B_w$, is
determined by weak P-odd and T-even interactions. According to
(\ref{eq182a}), corresponding terms in the scattering amplitude
can be written as:
\begin{equation}
\hat{F}_w (\vec{q})=(B_{we}(\vec{q}) + B_{w nuc}(\vec{q})) \vec{\sigma}\vec{N}_{w}.
\label{eq600}
\end{equation}
Contribution $B_{we}(\vec{q})$ caused by parity violating weak interaction between the baryon and electrons can be expressed as:

\begin{equation}
B_{we}(\vec{q})= \tilde{B}_{we}(\vec{q})\Phi_e(\vec{q}),
\label{eq601}
\end{equation}
where $\tilde{B}_{we}$ is the coefficient defining baryon elastic
scattering amplitude by resting electron
$\hat{f}_{we}(q)=\tilde{B}_{we}\vec{\sigma}\vec{N}_{w},
\Phi_e(\vec{q})=\int e^{-i \vec{q} \vec{r}} N_e (\vec{r})d^3r=Z$,
Z is the nucleus charge. Minor corrections caused by thermal
oscillations of atoms centers of gravity will not be considered
further. To take them into consideration one should multiply
$\Phi_e(\vec{q})$ by $\Phi_{osc}(\vec{q})$, which is the
form-factor defined by oscillations of atoms nucleus.

Term $B_{w nuc}(\vec{q})$ (see (\ref{eq600})), which caused by parity violating weak interaction between the baryon and nucleus can take the following form:
\begin{equation}
B_{w nuc}(\vec{q})= \tilde{B}_{w nuc}(\vec{q})\Phi_{osc}(\vec{q}),
\label{eq602}
\end{equation}
where $\tilde{B}_{w nuc}$ is the coefficient defining baryon elastic scattering amplitude by resting nucleus $\hat{f}_{w nuc}=\tilde{B}_{w nuc}\vec{\sigma}\vec{N}_w$.

Due to the short-range character of P-violating interactions coefficients $\tilde{B}_{we}(\vec{q})\simeq \tilde{B}_{we}(0)$ and $\tilde{B}_{w nuc}(\vec{q})\simeq \tilde{B}_{w nuc}(0)$ at angles $\vartheta \simeq \frac{\tau}{k}\ll 1 $. As a result following expressions can be obtained for the effective potential energy of interaction with a crystallic plane (axis), which is caused by P-violating interactions $ \hat{U}_w $
\begin{equation}
\hat{U}_w = \hat{U}_{we} + \hat{U}_{w nuc}
\label{eq603}
\end{equation}
a) for the case of plane:
\begin{eqnarray}
\hat{U}_{we}(x) &=& -\frac{2 \pi \hbar^2}{m \gamma dy dz} \tilde{B}_{we}(0) N_{e}(x) \vec{\sigma}\vec{N}_{w}, \nonumber \\
\hat{U}_{wnuc}(x) &=& -\frac{2 \pi \hbar^2}{m \gamma dy dz} \tilde{B}_{wnuc}(0) N_{nuc}(x) \vec{\sigma}\vec{N}_{w}, \nonumber \\
N_{e(nuc)}(x)&=& \int N_{e(nuc)} (x,y,z) dy dz. \label{eq604}
\end{eqnarray}
b) for the case of axis:
\begin{eqnarray}
\hat{U}_{we}(\vec{\rho}) &=& -\frac{2 \pi \hbar^2}{m \gamma dz} \left\{\tilde{B}_{we}(0) N_{e}(\vec{\rho}) + \tilde{B}_{wnuc}(0) N_{nuc}(\vec{\rho})\right\}\vec{\sigma}\vec{N}_{w}, \nonumber \\
N_{e(nuc)}(\vec{\rho})&=& \int N_{e(nuc)} (\vec{\rho}, z)dz.
\label{eq605}
\end{eqnarray}

Thus:

\begin{equation}
\hat{U}_{w}(x)= \hat{U}_{we}(x) + \hat{U}_{wnuc}(x)= - (\alpha_w (x) + i \delta_w (x)) \vec{\sigma}\vec{N}_{w},
\label{eq606}
\end{equation}
where
\begin{eqnarray}
\alpha_w (x)&=& \alpha_{we} (x)+\alpha_{wnuc} (x), \nonumber\\
 \delta_w (x)&=& \delta_{we} (x)+\delta_{wnuc} (x), \nonumber
\label{eq607}
\end{eqnarray}
\begin{eqnarray}
\alpha_w (x)&=& \frac{2 \pi \hbar^2}{m \gamma dy dz} (\tilde{B}'_{we}(0) N_{e}(x) + \tilde{B}'_{wnuc}(0) N_{nuc}(x)), \nonumber\\
\delta_w (x)&=& \frac{2 \pi \hbar^2}{m \gamma dy dz} (\tilde{B}''_{we}(0) N_{e}(x) + \tilde{B}''_{wnuc}(0) N_{nuc}(x)). \nonumber
\label{eq608}
\end{eqnarray}

\section{Effective potential energy $\hat{U}$ determined by the electric dipole moment and other T-nonivariant interactions}

Let us consider now the electric dipole moment and other
T-nonivariant contributions to the spin rotation. According to
(\ref{eq182a}), corresponding terms in the scattering amplitude
can be written as:
\begin{equation}
\hat{F}_T (q)=( B_{EDM}(q) + B_{Te}(q) + B_{T nuc}(q)) \vec{\sigma}\vec{q}.
\label{eq47}
\end{equation}
%
Lets consider the term $\hat{F}_{EDM} (q)=B_{EDM}(\vec{q})\vec{\sigma}\vec{q}$. The coefficient $B_{EDM}(q)$ has both real and imaginary parts $B_{EDM}(q)=B_{EDM}'+iB_{EDM}''$.
By the approach used for deriving $\hat{F}_{magn} (q)$, for $\hat{F}_{EDM }(\vec{q})$ we obtain:

\begin{eqnarray}
\hat{F}_{EDM} (\vec{q})&=& -i\frac{m \gamma d}{2 \pi \hbar^2} V_{coul}(\vec{q})\vec{\sigma}\vec{q}+ \nonumber\\
&+&\frac{k}{4 \pi \hbar^2 c^2} \iint e^{-i \vec{q}_{\perp} \vec{r}_{\perp} }\left\{\overline{\left[\int \hat{V} (\vec{r}_{\perp},z)dz\right]^2}-\left[\int \overline{ \hat{V} (\vec{r}_{\perp},z)}dz\right]^2\right\}d^2 r_{\perp},\nonumber\\
\label{eq48}
\end{eqnarray}
where $\hat{V}(\vec{r})=V_{coul}(\vec{r})+V_{EDM}(\vec{r}) $,
$V_{EDM}=-D\vec{\sigma}\vec{E}$ is the energy of interaction
between the electric dipole moment $D$ and the electric field
$\vec{E}$,  $D=ed$,  $e$ is the electric charge of the particle.

Using (\ref{eq25}), the following expression for the potential energy of interaction between the particle and the plane can be obtained:
\begin{equation}
\hat{U}_{EDM} = -ed E_{pl}(x)\vec{\sigma}\vec{N}_{T}- i\frac{d}{2 d_y d_z \hbar c}\frac{\partial}{\partial x} \overline{\delta V^2 (x)} \vec{\sigma}\vec{N}_{T},
\label{eq49}
\end{equation}
where the unit vector $\vec{N}_T$ is orthogonal to the plane, $\vec{E}_{pl}(x)=E_x \vec{N}_{T}$.

Evidently, $\hat{U}_{EDM}$ can be written as:

\begin{equation}
\hat{U}_{EDM} = -(\alpha_{EDM} +i\delta_{EDM}) \vec{\sigma}\vec{N}_{T}.
\label{eq50}
\end{equation}
Similar to $\hat{U}_{magn}$, the energy $\hat{U}_{EDM}$ has both real and imaginary parts.
The expression for  $\hat{U}_{magn}$ converts to $\hat{U}_{EDM}$
when replacing $ \frac{g-2}{2} \rightarrow 2\frac{d}{\lambda_c}$
($ \lambda_c=\frac{\hbar}{mc} $ is the Compton wave-length of the
particle) and $\vec{N} \rightarrow \vec{N}_T  $. Therefore:
\begin{equation}
\frac{U_{EDM}}{U_{magn}}=\frac{4d}{\lambda_c (g-2)}=\frac{ed}{\mu_A}=\frac{D}{\mu_A},
\label{eq51}
\end{equation}
%
As is expected, the above expression is equal to the ratio of electric dipole moment to anomalous magnetic moment.

Lets remind that amplitude $\hat{F}_{T} (\vec{q})$ contains both terms caused by EDM and those determined by short-range T-noninvariant interactions between the baryon and electrons and nuclei $B_{T e}(\vec{q})$ and $B_{T nuc} (q)$. Contribution caused by these terms should also be added to the effective potential energy of the interaction between baryon spin and nuclei of the crystal $\hat{U}_{T} (x)$:

\begin{equation}
\hat{U}_{T} (x)= \hat{U}_{EDM} + \hat{U}_{Te} + \hat{U}_{T nuc} = -(\alpha_{T}(x) + i\delta_{T}(x))\vec{\sigma}\vec{N}_{T},
\label{eq52}
\end{equation}
where $\alpha_{T} =\alpha_{EDM}+ \alpha_{Te}+ \alpha_{T nuc}, \delta_{T}=\delta_{EDM}+ \delta_{Te}+ \delta_{T nuc}$.

Expressions for coefficients  $\alpha_{Te(nuc)}$ and
$\delta_{Te(nuc)}$ can be evaluated in terms of scattering
amplitude by the following way. Lets define the form-factor
determined by electrons distributions in atom and nucleus
oscillations.
\begin{equation}
B_{Te} (\vec{q})=\tilde{B}_{Te}(\vec{q})\Phi_{e}(\vec{q}),
B_{Tnuc} (\vec{q})=\tilde{B}_{Tnuc}(\vec{q})\Phi_{osc}(\vec{q}),
\label{eq53}
\end{equation}
where $\Phi_{e}(\vec{q})=\int e^{-i \vec{q} \vec{r}} N_{e}
(\vec{r}) d^3 r$, $ N_{e} (\vec{r}) $  is electrons density
distribution in atom, $\int N_{e}(\vec{r})d^3 r=Z$, Z is the
nucleus charge, $\Phi_{osc}(\vec{q})$ is determined by
(\ref{eq30}), $\tilde{B}_{Te}$ is the coefficient defining baryons
scattering amplitude by resting electron
$\hat{f}_{Te}=\tilde{B}_{Te}(\vec{q})\vec{\sigma}\vec{q} $,
$ \tilde{B}_{nuc} (q)$  is the coefficient defining baryons
scattering amplitude by resting nucleus
$\hat{f}_{Tnuc}=\tilde{B}_{nuc}(\vec{q})\vec{\sigma}\vec{q}$.
%
Lets remind that in compliance with (\ref{eq12}) the contribution caused by elastic coherent scattering should be subtracted from the amplitude $B_{T}$.

Due to the short-range character of T-noninvariant interactions
coefficients $\tilde{B}_{Te}(\vec{q})\simeq\tilde{B}{_{Te}}(0)$
and $\tilde{B}_{nuc}(\vec{q})\simeq\tilde{B}_{nuc}(0)$ at angles
$\vartheta\simeq\frac{\tau}{k}<<1$. As a result following
expressions can be obtained:
\begin{eqnarray}
&\hat{U}_{Te}(x)&=i \frac{2 \pi \hbar^2}{m \gamma d_y d_z} \tilde{B}_{Te} (0) \frac{dN_{e} (x)}{dx} \vec{\sigma}\vec{N}_{T},\nonumber\\
&\hat{U}_{Tnuc}(x)&=i \frac{2 \pi \hbar^2}{m \gamma d_y d_z} \tilde{B}_{Tnuc} (0) \frac{dN_{nuc} (x)}{dx} \vec{\sigma}\vec{N}_{T},\nonumber\\
&N_{e(nuc)}(x)&=\int N_{e(nuc)}(x,y,z)dydz,
\label{eq54}
\end{eqnarray}
Coefficients $\tilde{B}_{Te} (0)$ and $\tilde{B}_{Tnuc} (0)$ are complex values:
\begin{equation}
\tilde{B}_{Te(nuc)}(0)=\tilde{B}'_{Te(nuc)}+i\tilde{B}''_{Te(nuc)}.\nonumber
\end{equation}

As a result we have:
\begin{equation}
\hat{U}_{Te(nuc)}(x)=-(\alpha_{Te(nuc)}+i\delta_{Te(nuc)})\vec{\sigma}\vec{N}_{T},
\label{eq55}
\end{equation}
where
\begin{equation}
\alpha_{Te(nuc)}=\frac{2 \pi \hbar^2}{m \gamma dy dz} \tilde{B}''_{Te(nuc)} \frac{dN_{e(nuc)}(x)}{dx},\nonumber\\
\delta_{Te(nuc)}=\frac{2 \pi \hbar^2}{m \gamma dy dz} \tilde{B}'_{Te(nuc)} \frac{dN_{e(nuc)}(x)}{dx}.\nonumber
\end{equation}
Thus in the experiment aimed to obtain the limit for the EDM value, the limits for the scattering amplitude, which is determined by T(CP)-noninvariant interactions between baryons and electrons, and nuclei, will be found as well.
%
The obtained values of these amplitudes for different interaction
types allows one to restore the values of corresponding constants,
too. The simplest model for such a potential is the Yukawa
potential \cite{b28}.
Using it all the equations for $\alpha_{Te(nuc)}$ can be obtained
by replacement in (\ref{eq37}) of $ V_{coul}+V_{EDM} $ by
$V_{coul}+V_{T}$, $V_{T}=-d_{T} \vec{\sigma} \vec{r} \frac{e^
{-\varkappa_{T}r}} {r^2}$.
Here $d_{T}$ is the interaction constant, $\varkappa_{T} \sim
\frac{1}{M_{T}} $, where $ M_{T} $ is the mass of heavy particles,
exchange of which leads to  interaction $V_{T}$ \cite{b28}.
It should be noted that constant $d_{T}$ for interaction between
the heavy baryon and the nucleon can be greater than that for
nucleon-nucleon interaction.
This effect can be explained by the reasoning similar to that
explaining expected EDM growth for the heavy baryon.
T-odd interaction mixes baryon stationary states with different
parity more effectively that occurs due to probably smaller
spacing between energy levels corresponding to these states.

\section{P and CP violating spin rotation in bent crystals}
Expressions for the energy of interaction between the baryon and
the plane(axis) that were obtained above, allow us to find the
equation describing evolution of the particle polarization vector
in a bent crystal. The mentioned equations differ from those,
which describe evolution of the spin in external electromagnetic
fields in vacuum,  by the presence of terms defining the
contribution of P and T(CP) noninvariant interactions between
electrons and nuclei to the spin rotation.

These equations can be obtained by the following approach \cite{b12}. The spin wave function $|\Psi(t)>$ meets the equation as follows: 

\begin{equation}
ih \frac{\partial |\Psi(t)>}{\partial t}= \hat{U}_{eff}|\Psi(t)>.
\label{eq56}
\end{equation}
The baryon polarization vector $\vec{\xi}$ can be found via $|\Psi(t)>$:

\begin{equation}
\vec{\xi}=\frac{<\Psi(t)|\vec{\sigma}|\Psi(t)> }{<\Psi(t)|\Psi(t)>},
\label{eq57}
\end{equation}
Thus the equation for spin rotation of a particle $(\gamma >>1)$, which moves in a bent crystal, reads as follows:

\begin{eqnarray}
\frac{d \vec{\xi}}{dt}&=&-\frac{e(g-2)}{2mc}[\vec{\xi}\times[\vec{n}\times\vec{E}]]
-\frac{2}{\hbar}\delta_{m}\{{ \vec{N}_{m}-\vec{\xi}(\vec{N}_{m}\vec{\xi})}\}-\nonumber\\
&-&\frac{2}{\hbar}\alpha_{s0}[\vec{\xi}\times\vec{N}_{m}]-
\frac{2}{\hbar}\delta_{s0}
\{{\vec{N}_{m}-\vec{\xi}(\vec{N}_{m}\vec{\xi})}\} +\nonumber\\
&+&\frac{2ed}{\hbar}[\vec{\xi}\times\vec{E}]
+\frac{2}{\hbar}\delta_{EDM}\{{\vec{N}_{T}-\vec{\xi}(\vec{N}_{T}\vec{\xi})}\}
+\nonumber\\
&+&\frac{2}{\hbar}(\alpha_{Te}+\alpha_{Tnuc})[\vec{\xi}\times
\vec{N}_{T}] +
\frac{2}{\hbar}(\delta_{Te} + \delta_{Tnuc})\{\vec{N}_{T} - \vec{\xi}(\vec{N}_{T}\vec{\xi})\} + \nonumber\\
&+& \frac{2}{\hbar}\alpha_{w}[\vec{\xi}\times \vec{n}] -
\frac{2}{\hbar}\delta_{w} \{\vec{n}-\vec{\xi}(\vec{\xi}\vec{n})\}.
\label{eq58}
\end{eqnarray}
Let us note that vector $[\vec{n} \times \vec{E}]$ is parallel to
vector $\vec{N}_{m}=[\vec{n} \times \vec{n}_{x}]$, vector
$\vec{E}$ is parallel to $\vec{N}_{T}=\vec{n}_{x}$, $
\vec{n}=\frac{\vec{k}}{k} $ is the unit vector parallel to the
direction of the particles momentum. Equation (\ref{eq58}) can be
also expressed as:

\begin{eqnarray}
\frac{d \vec{\xi}}{dt}&=& - \left(\frac{e(g-2)}{2mc} E_{x} (x) + \frac{2}{\hbar}\alpha_{s0}(x)\right)[\vec{\xi}\times\vec{N}_{m}]- \nonumber\\
&-& \frac{2}{\hbar}\left.\bigg(\delta_{m}(x)+\delta_{s0}(x)\right.\bigg) \{\vec{N}_{m}-\vec{\xi}(\vec{N}_{m}\vec{\xi})\}+ \nonumber\\
&+&\frac{2}{\hbar} \left.\bigg( edE_{x} (x) + \alpha_{Te} (x) + \alpha_{Tnuc} (x)\right.\bigg) [\vec{\xi}\times\vec{N}_{T}]+\nonumber\\
&+&\frac{2}{\hbar}\left.\bigg(\delta_{EDM} (x) +\delta_{Te}(x)+\delta_{Tnuc}(x)\right.\bigg)\{\vec{N}_{T}-\vec{\xi}(\vec{N}_{T}\vec{\xi})\}+ \nonumber\\
&+&\frac{2}{\hbar}\alpha_{w}[\vec{\xi}\times\vec{n}]-\frac{2}{\hbar}
\delta_{w}\{\vec{n}-\vec{\xi}(\vec{\xi}\vec{n})\}. \label{eq**}
\end{eqnarray}
According to (\ref{eq**}) baryon spin rotates about three axes
\cite{b29}: the effective magnetic field direction
$\vec{N}_{m}||[\vec{n}\times\vec{E}]$, the electric field
direction $\vec{N}_{T}||\vec{E}$ and the direction of the momentum
$\vec{n}$.

Nonelastic processes in crystals cause appearance of different types of contributions in (\ref{eq58},\ref{eq**}): terms proportional to $\delta$ lead to rotation of the polarization vector in directions of vectors  $\vec{N}_{m}$, $\vec{N}_{T}$ and $\vec{n}$. As is also seen from (\ref{eq**}), when an unpolarized beam enters a crystal polarization in direction of vectors  $\vec{N}_{m}$, $\vec{N}_{T}$ and $\vec{n}$ arises \cite{b29}.

Contributions to the equation (\ref{eq**}), which are caused by the interaction between baryon and nuclei, depend on distribution of nuclei density  $N_{nuc}(x)$ (see terms proportional to  $\alpha_{s0}(x), \delta_{s0}(x), \alpha_{Tnuc}(x), \delta_{Tnuc}(x)$). As a result, for positively charged particles, moving in the channel along the trajectories located in the center of the channel, such contributions are suppressed.
%

Thus, according to (\ref{eq**}), when conducting and interpreting
the experiments for measuring EDM, one should take into
consideration the fact  that  measuring spin rotation provides
information about the sum of contributions to T-noninvariant
rotation. The stated rotation is determined  by both EDM and
short-range CP-noninvariant interactions. Nonelastic
T-noninvariant processes lead to spin rotation in direction of $
\vec{N}_{T}$ as well, which gives additional opportunities for EDM
measurement.
%

Lets evaluate the most important new effects described by the
equation(\ref{eq**}) and consider the contribution to spin
rotation caused by spin rotation in direction of  $ \vec{N}_{m} $.
According to (\ref{eq41}) coefficient  $ \delta_{m} $ has the
following form:
\begin{eqnarray}
\delta_{m}&=& \frac{1}{4d_yd_zmc^2} (\frac{g-2}{2}) \frac{\partial}{\partial x} \overline{\delta V^2 (x)}= \nonumber \\
&=&\frac{1}{4d_yd_z}mc^2(\frac{g-2}{2}) \frac{\partial}{\partial x} \int \left\{ \overline{\left[\int V_{coul} (x,y,z)dz\right]^2} - \right. \nonumber \\
&-& \left. \left[\int \overline{V_{coul} (x,y,z)}dz\right]^2\right\}dy,
\label{eq59}
\end{eqnarray}
where $V_{coul} (x,y,z)=\sum_i V_e (x-x_i, y-y_i,z-z_i) - V_{nuc}
(x-\eta_{fx},y-\eta_{fy}, z-\eta_{fz})$, $x_{i},y_{i},z_{i}$ are
the coordinates of the $i$-th electron in atom, $
\eta_{fx},\eta_{fy}, \eta_{fz} $ are the coordinates of nucleus.
Let us choose the equilibrium point position of the  oscillating
nucleus as the origin of coordinates. The line denotes averaging
of electrons and nuclei positions by electrons density
distribution and nuclei oscillations; in other words, averaging
with wave-functions of atoms in crystal. By means of these
functions, the density distribution takes the form:

\begin{equation}
N(\vec{r}_1, \vec{r}_2......\vec{r}_z,\vec{\eta})=N_e (\vec{r}_1, \vec{r}_2......\vec{r}_z,\vec{\eta}) N_{nuc} (\vec{\eta}),
\label{eq60}
\end{equation}
where $N_{e}$ is the density distribution of electrons in atom, $N_{nuc}(\vec{\eta})$ is the density distribution of nucleus oscillations.

Lets introduce the function $W(x,y) = \int V(x,y,z)dz $. From (\ref{eq59}) we have:

\begin{eqnarray}
W(x,y)&=&\sum_i \int V_e (x-x_i, y-y_i,\xi) d\xi - \int V_{nuc} (x-\eta_x, y-\eta_y,\xi)d\xi= \nonumber \\
&=& \sum_i W_e (x-x_i, y-y_i)- W_{nuc}(x-\eta_x, y-\eta_y)
\label{eq61}
\end{eqnarray}

\begin{eqnarray}
\overline{W^2 (x,y)} =\int [\sum_i W_e (\vec{\rho}-\vec{\rho}_i) - W_{nuc} (\vec{\rho}-\vec{\eta}_{\perp})]^2 \times \nonumber \\
\times N_e (\vec{\rho}_1-\vec{\eta}_1,........\vec{\rho}_z-\vec{\eta}_{\perp}) N_{nuc}(\vec{\eta}_{\perp}) d^2\rho_1 d^2\rho_z d^2 \vec{\eta}_{\perp},
\label{eq62}
\end{eqnarray}
where $\vec{\rho}=(x,y),\vec{\eta}_{\perp} =(\eta_{x}, \eta_{y}), Z $ is the number of electrons in atom.

In other words:
\begin{eqnarray}
& &\overline{W^2 (\vec{\rho})} =\int \{ (\sum_i W_e (\vec{\rho}-\vec{\rho}_i))^2-2\sum W_e (\vec{\rho}-\vec{\rho}_i) W_{nuc} (\vec{\rho}-\vec{\eta}_{\perp}) + \nonumber \\
&+& W^2_{nuc} (\vec{\rho}-\vec{\eta}_{\perp})\}N_e (\vec{\rho}_1-\vec{\eta}_1,........\vec{\rho}_z-\vec{\eta}_{\perp}) N_{nuc}(\vec{\eta}_{\perp}) d^2\rho_1 d^2\rho_z d^2 \vec{\eta}_{\perp}. \nonumber \\
\label{eq63}
\end{eqnarray}

Averaging $\overline{W^{2}(\rho)}$, provides appearing the
expressions for both density distribution of a single electron in
atom and those dependent on coordinates of two electrons in the
atom, which describe pair correlations in electrons positions in
atom. However, the influence of pair correlations will be ignored
during the estimations. As a result the expression (\ref{eq63})
can be represented as follows:

\begin{eqnarray}
\overline{W^2 (\rho)} &=&\int d^2 \eta_{\perp} \{Z [\langle W_e^2(\vec{\rho},\vec{\eta}_{\perp}) \rangle _e^2 -\langle W_e (\rho,\eta_{\perp})\rangle _e^2] + Z^2 <W_e(\vec{\rho},\vec{\eta}_{\perp})> _e^2 - \nonumber \\
&-& 2Z <W_e(\vec{\rho},\vec{\eta}_{\perp})> W_{nuc}(\vec{\rho}-\vec{\eta}_{\perp}) +  W^2_{nuc}(\vec{\rho}-\vec{\eta}_{\perp}) \},
\label{eq64}
\end{eqnarray}
where the function
\begin{eqnarray}
<W_e(\vec{\rho},\vec{\eta}_{\perp})>_e= \int W_e (\vec{\rho}-\vec{\rho} ') N_e (\vec{\rho} '-\vec{\eta}_{\perp})d^2 \rho ' \nonumber\\
<W_e^2 (\rho,\vec{\eta}_{\perp} )>_e= \int W_e^2(\vec{\rho}-\vec{\rho}')N_e (\vec{\rho} '-\vec{\eta}_{\perp})d^2 \rho',\nonumber
\end{eqnarray}
that means

\begin{eqnarray}
\overline{W^2(\vec{\rho})}= \int d^2 \eta_{\perp} \{ Z [<W_e^2(\vec{\rho},\vec{\eta}_{\perp})>_e -<W_e(\vec{\rho},\eta_{\perp})>^2_e] + \nonumber \\
+ (Z<W_e(\vec{\rho},\vec{\eta}_{\perp})>_e -W_{nuc}(\vec{\rho}-\vec{\eta}_{\perp}))^2 \}N_{nuc}(\vec{\eta}_{\perp}).
\label{eq66}
\end{eqnarray}

According to (\ref{eq59}), the function $\int
\left[\overline{W^2(\vec{\rho})} -
\overline{W(\vec{\rho})}^2\right]dy$ determines the expression for
$ \delta_m $. It should be noted that when fluctuations caused by
nuclei oscillations are neglected, only fluctuations, which are
determined by distribution of electrons coordinates in atom, are
left.

As a result, the following equation for $ \delta_m $ can be obtained:
\begin{equation}
\delta_m= \frac{1}{4 d_y d_z} mc^2  \left(\frac{g-2}{2}\right) \frac{\partial}{\partial x} \int \left\{ \overline{W^2 (x,y)} - \overline{W (x,y)}^2 \right\} dy,
\label{eq67}
\end{equation}
where $ \overline{W(\vec{\rho})}= \int \left\{Z \overline{W_e (\vec{\rho}, \eta_{\perp})}^e -W_{nuc} (\vec{\rho}- \eta_{\perp}) \right\}N_{nuc} (\eta_{\perp})d^2 \eta_{\perp}$. $\overline{W_e (\vec{\rho}, \eta_{\perp})}^e = \int W_e (\vec{\rho}-\vec{\rho} ')N_e (\vec{\rho} '-\vec{\eta_{\perp}})d^2 \rho ' $.

The following estimation for the value $ \delta_m $ can be
obtained from (\ref{eq67}): $ \frac{1}{\hbar}\delta_m \sim
10^{8}\div10^{9} sec^{-1}$. According to \cite{b10,bn11} the charm
baryon EDM can be as large as $ d \sim 10^{-17}$. Spin rotation
frequency $\Omega_{EDM}$ determined by such charmed baryon EDM is
$\Omega_{EDM} \sim 10^{6}-10^{7}sec^{-1}$. As a result, nonelastic
processes caused by magnetic moment scattering  can imitate the
EDM contribution.

The contributions of P-odd and T-even rotation effect to the general spin rotation can be evaluated by the following way.
Precession frequency $\Omega_{w}=\frac{2}{\hbar}\alpha_w$ is
determined by the real part of the amplitude of baryon weak
scattering by an electron (nucleus). This amplitude can be
evaluated in the energy range of about W and Z bosons production
and smaller by Fermi theory \cite{b26+}:
\begin{equation}
\label{eq29.1}
ReB\sim G_{F}k=10^{-5}\frac{1}{m^{2}_{p}}k=10^{-5}\frac{\hbar }{m_{p} c}\frac{m}{m_{p}\gamma}=10^{-5}\lambda_{cp} \frac{m}{m_{p}\gamma} ,
\end{equation}
where $ G_{F} $ is the Fermi constant, $m_{p}$ is the proton mass,
$\lambda_{cp} $ is the proton Compton wavelength. For particles
with energy from hundreds of GeV to TeV $ReB\sim G_{F}k=10^{-16}$
cm.

For different particle trajectories in a bent crystal precession
frequency $ \Omega_{w} $ could vary in the range $
\Omega_{w}\simeq 10^{3}\div 10^{4} sec^{-1}$. Therefore, when
particle a passes 10~cm in a crystal, its spin undergoes
additional rotation around momentum direction at angle $
\vartheta_{p} \simeq 10^{-6}\div 10^{-7}$ rad.
The effect grows for a heavy baryon as a result of the mechanism
similar to that of its EDM growth (see the explanation for the
growth of constant $d_T$ mentioned above).

Absorption caused by parity violation weak interaction also
contributes to change of spin direction (see in
(\ref{eq58},\ref{eq**}) the terms proportional to $\delta_w$).
This rotation is caused by imaginary part of weak scattering
amplitude and is proportional to the difference of total
scattering cross-sections  $ \sigma_{\uparrow\uparrow} $ and $
\sigma_{\downarrow\uparrow} $ \cite{b29}.

This difference is proportional to the factor, which is determined
by interference of coulomb and weak interactions for baryon
scattering by an electron, and  of strong (coulomb) and weak
interactions for baryon scattering by nuclei \cite{b29}.
\begin{equation}
\sigma_{\uparrow\uparrow (\downarrow\uparrow)}=\int| f_{c(nuc)}+B_{0w}\pm B_{w}|^{2} d\Omega ,
\label{eq30.1}
\end{equation}

\begin{equation}
\sigma_{\uparrow\uparrow} - \sigma_{\downarrow\uparrow}= 2\int[(f_{c(nuc)}+B_{0w})B^{*}+( f_{c(nuc)}+B_{0w})^{*}B]d\Omega .
\label{eq31.1}
\end{equation}

When baryon trajectory passes in the area, where collisions with
nuclei are important (this occurs in the vicinity of potential
barrier for positively charged particles),  the value $
\frac{2}{\hbar}\delta_w \sim 10^{6}\div10^{7} sec^{-1}$.
Similar to the real part $ReB$ for the case of heavy baryons the
difference in cross-sections grows.
 Multiple scattering also
contributes to spin rotation \cite{b12,b22}. Particularly,  due to
interference of weak and coulomb interactions the root-mean-square
scattering angle appears changed and dependent on spin orientation
with respect to the particle momentum direction \cite{b29}.

When measuring MDM and T-odd spin rotation in a bent crystal, one
can eliminate parity violating rotation by the following way. MDM
and T-odd spin rotations, unlike P-odd spin rotations, depend on
crystal turning at $ 180^{\circ} $ around the direction of
incident baryon momentum. Namely, P-odd effect does not change,
while the sign of MDM and T-odd spin rotations does due to change
of the electric field direction. Subtracting results of
measurements for two opposite crystal positions on could obtain
the angle of rotation, which does not depend on P-odd effect.

\section{Conclusion}
Besides electromagnetic interaction the channelled particle moving
in a crystal experiences weak interaction with electrons and
nuclei as well as strong interaction with nuclei. Mentioned
interactions lead to the fact that in the analysis of the
particle's spin rotation, which is caused by  electric dipole
moment interaction with electric field, both $P_{odd}, T_{even}$
and $P_{odd}, T_{odd}$ non-invariant spin rotation resulting from
weak interaction should be considered.
As obtained here, spin precession of channelled particles in bent
crystals at the LHC gives unique possibility for measurements of
both electric and magnetic moments of charm, beauty and strange
charged baryons, as well as constants determining  CP ($T_{odd},
P_{odd}$) violation interactions and $P_{odd}, T_{even}$
interactions of baryons with electrons and nucleus (nucleons). For
a particle moving in a bent crystal a new effect caused by
nonelastic processes arises: in addition to the spin precession
around the direction of the effective magnetic field (bend axis),
the direction of electric field and the direction of the particle
momentum, the spin rotation to the mentioned directions also
appears.

\section{Annex}
Let us  consider scattering of a relativistic particle in crystal,
formed by the set of $ N $ atoms. It should be reminded, that the
Dirac equation describing scattering process can be transformed to
the Schr\"{o}dinger equation for the particle with relativistic
mass $M=m\gamma$, where $m$ is the rest mass of the particle,
$\gamma$ is the Lorentz factor.
The corresponding equation is
\begin{eqnarray}
\label{4.31}
\left(E_{a}-H(\xi_{1}\ldots\xi_{N}) +\frac{\hbar^{2}}{2m\gamma}\Delta_{r}\right) \psi(\vec{r},\xi_{1}\ldots\xi_{N}) \nonumber \\
=\sum^{N}_{i=1} V_{i}(\vec{r},\xi_{i}) \psi(\vec{r},\xi_{1}\ldots\xi_{N})\,,
\end{eqnarray}
where $H(\xi_{1}\ldots\xi_{N})$ is the  Hamiltonian of the scatters; $\xi_{1}\ldots\xi_{N}$
is the set of coordinates describing the first and other  scatterers ($\xi$ also includes spin variables);
$V_{i}(\vec{r}, \xi_{i})$ is the energy of the interaction between the incident particle and the $i$-th scatterer;
$\vec{r}$ is the coordinate of the incident particle.
If $G(\vec{r},\xi_{1}\ldots\xi_{N};\,\vec{r}^{\,\prime},\xi_{1}^{\prime}\ldots\xi_{N}^{\prime})$
is the Green function of the operator
\[ E_{a}-H+\frac{\hbar^2}{2m\gamma}\Delta_{r}\,
\]
then (\ref{4.31}) can be written in the form

\begin{eqnarray}
\label{4.32}
\psi_a(\vec{r},\xi_1\ldots\xi_N)&=&\Phi_a(\vec{r},\xi_1\ldots\xi_N) \nonumber \\
&+&\int\int G(\vec{r},\xi_{1}\ldots\xi_{N};\vec{r}^{\,\prime},\xi_{1}^{\prime}\ldots\xi_{N}^{\prime}) \sum^N_{i=1} V_i(\vec{r}^{\,\prime},\xi_i^{\prime}) \nonumber \\
&\times&\psi_a(\vec{r}^{\,\prime},\xi_1^{\prime}\ldots\xi_N^{\prime})d_3 r^{\prime}d_3\xi_1^{\prime}\ldots d_3\xi_N^{\prime}
\end{eqnarray}
[ $\Phi_a$ are the eigenfunctions of the operator $\left(-\frac{\hbar^2}{2m\gamma}\Delta_r+H\right)$].
Taking into account that \cite{b12}

\begin{equation}
\label{4.33}
\sum_{i}V_i(\vec{r},\xi_i)\psi_a(\vec{r},\xi_1\ldots\xi_N)=T(\vec{r},\xi_1\ldots\xi_N)\Phi_a(\vec{r},\xi_1\ldots\xi_N)\,,
\end{equation}
where $T$ is the  operator of scattering by $N$ centers, the following equation can be derived for $T$:

\begin{eqnarray}
\label{4.34}
& &T(\vec{r},\xi_1\ldots\xi_N)\Phi_a(\vec{r},\xi_1\ldots\xi_N)= \sum^N_{i=1}V_i(\vec{r},\xi_i)\Phi_a(\vec{r},\xi_1\ldots\xi_N) \nonumber \\
& &+\sum^N_{i=1}V_i(\vec{r},\xi_i)\int\int G(\vec{r},\xi_1\ldots\xi_N;\vec{\rho},\vec{\eta}_1\ldots\vec{\eta}_N) \nonumber \\
& &\times  T(\vec{\rho},\vec{\eta}_1\ldots\vec{\eta}_N)\Phi_a(\vec{\rho},\vec{\eta}_1\ldots\vec{\eta}_N)d^3\rho d^3\eta_1 \ldots d^\eta_N \,.
\end{eqnarray}
Let us introduce the notation
$T(\vec{r},\xi_1\ldots\xi_N)\Phi_a(\vec{r},\xi_1\ldots\xi_N)=T_a(\vec{r},\xi_1\ldots\xi_N)$.
Then, it is convenient to introduce the operators $T^{i}$, using
the equalities:

\begin{eqnarray}
\label{4.35}
T^{(i)}_a(\vec{r},\xi_1\ldots\xi_N)= V_i(\vec{r},\xi_i)\Phi_a(\vec{r},\xi_1\ldots\xi_N) \nonumber \\
+V_i(\vec{r},\xi_i)\int\int G(\vec{r},\xi_1\ldots\xi_N;\,\vec{\rho},\vec{\eta}_1\ldots\vec{\eta}_N)  \nonumber \\
\times T_a(\vec{\rho},\vec{\eta}_1\ldots\vec{\eta}_N)d^3\,\rho d^3\eta_1 \ldots d^3\eta_N\,,
\end{eqnarray}
i.e., $T=\sum\limits_{i}T^{(i)}$.
The system (\ref{4.34}) can be represented as

\begin{eqnarray}
\label{4.36}
T^{(i)}_a(\vec{r},\xi_1\ldots\xi_N)= t^{(i)}_a(\vec{r},\xi_1\ldots\xi_N) \nonumber \\
+t^{(i)}(\vec{r},\xi_1\ldots\xi_N)\int\int G(\vec{r},\xi_1\ldots\xi_N;\,\vec{\rho},\vec{\eta}_1\ldots\vec{\eta}_N) \nonumber \\
\times \sum_{l\neq i}T_a^{(l)}(\vec{\rho},\vec{\eta}_1\ldots\vec{\eta}_N) d^3\rho d^3\eta_1 \ldots d^3\eta_N \,,
\end{eqnarray}
where

\begin{eqnarray*}
    t^{(i)}_a(\vec{r},\xi_1\ldots\xi_N)= V_i(\vec{r},\xi_i)\Phi_a(\vec{r},\xi_1\ldots\xi_N)  \\
    +V_i(\vec{r},\xi_i)\int \int G(\vec{r},\xi_1\ldots\xi_N;\,\vec{\rho},\vec{\eta}_1\ldots\vec{\eta}_N) \\
    \times t_a^{(i)}(\vec{\rho},\vec{\eta}_1\ldots\vec{\eta}_N)d^3\rho d^3\eta_1 \ldots d^3\eta_N\,.
\end{eqnarray*}
As is known \cite{bn13,bn15},
\begin{eqnarray*}
    G(\vec{r},\xi_1\ldots\xi_N;\,\vec{\rho},\vec{\eta}_1\ldots\vec{\eta}_N)  \\
    =-\frac{m\gamma}{2\pi \hbar^2}\sum_b
    \varphi_b(\xi_1\ldots\xi_N)\varphi_b^{\ast}(\vec{\eta}_1\ldots\vec{\eta}_N)
    \frac{e^{ik_b|\vec{r}-\vec{\rho}|}}{|\vec{r}-\vec{\rho}|}\,,
\end{eqnarray*}
where $\varphi_b(\xi_1\ldots\xi_N)$ are the eigenfunctions of the operator $H(\xi_1\ldots\xi_N)$;
\[
k_b^2\equiv \frac{2m\gamma}{\hbar^2} \left(
E_A+\frac{\hbar^2 k_a^2}{2m\gamma}-E_B \right)=\frac{2m\gamma}{\hbar^2}(E_a-E_B)\,;
\]
$E_A$ and $E_B$ are the internal energies of the scattering system before and after the collision, respectively.

If the scatterers are independent of each other, the wave function $\varphi_{b}(\xi_{1}\ldots\xi_{N})$ is represented as the product of the wave functions of the scatterers:
\[
\varphi_{b}(\xi_{1}\ldots\xi_{N}) =\prod \varphi_{bi}(\xi_{i})\,.
\]
In this case the direct substitution can verify that
$t^{(i)}(\vec{r},\xi_{1}\ldots\xi_{N})=t^{(i)}(\vec{r},\xi_{i})$,
where $t^{(i)}(\vec{r},\xi_{i})$ is the operator of particle
scattering by the $i$-th  center in the absence of other centers.

Note now that the quantities $T^{(i)}_{a}$ can be written as follows:
\begin{equation}
\label{4.37}
T^{(i)}_{a}=t^{(i)}(\vec{r},\xi_{1}\ldots\xi_{N})\mathscr{F}^{(i)}_{a}(\vec{r},\xi_{1}\ldots\xi_{N})\,,
\end{equation}
where
\begin{eqnarray}
\label{4.38}
\mathscr{F}^{(i)}_{a}(\vec{r},\xi_{1}\ldots\xi_{N})&=&\Phi_{a}(\vec{r},\xi_{1}\ldots\xi_{N})+ \int\int G(\vec{r},\xi_{1}\ldots\xi_{N};\,\vec{\rho},\vec{\eta}_{1}\ldots\vec{\eta}_{N})  \nonumber \\
&\times&\sum_{l\neq i} T_{a}^{(l)}(\vec{\rho},\vec{\eta}_{1}\ldots\vec{\eta}_{N})d^{3}\rho d^{3}\eta_{1}\ldots d^{3}\eta_{N}\,.
\end{eqnarray}
The first term in (\ref{4.38}), being the function of $\vec{r}$,
describes the initial wave falling upon the $i$-th scatterer. The
second term can be interpreted as the contribution to the wave incident on
the $i$-th center that is due to scattering by other centers. Indeed,
if the interaction of the incident particle with all the centers
excepting for the $i$-th center  equaled  zero, then the second term would also equal  zero.

Let us make use of the definition (\ref{4.33}) and rewrite equation (\ref{4.32} )for the wave function $\psi_{a}$ in the form:
\begin{eqnarray}
\label{4.39}
\psi_{a}(\vec{r},\xi\ldots\xi_{N})&=&\Phi_{a}(\vec{r},\xi_{1}\ldots\xi_{N}) \nonumber \\
&+&\int\int G(\vec{r},\xi_{1}\ldots\xi_{N}; \vec{r}^{\,\prime},\xi_{1}^{\prime}\ldots\xi_{N}^{\prime} )\sum^{N}_{i=1}t^{(i)}(\vec{r}^{\,\prime},\xi_{1}^{\prime}\ldots\xi_{N}^{\prime}) \nonumber \\
&\times&\mathscr{F}_{a}^{(i)}(\vec{r}^{\,\prime},\xi_{1}^{\prime}\ldots\xi_{N}^{\prime})d^{3}r^{\prime}d^{3}\xi_{1}^{\prime}\ldots d^{3}\xi_{N}^{\prime}\,.
\end{eqnarray}
From (\ref{4.39}) follows that the  probability amplitude $\psi_{ba}(\vec{r})=\langle\varphi_b|\psi_a\rangle$ to find
the particle at point $\vec{r}$ and the system in state $b$
satisfies the system of equations

\begin{eqnarray}
\label{4.40}
\psi_{ba}(\vec{r})=e^{ik_a \vec{r}}\delta_{ba}-\frac{m\gamma}{2\pi \hbar^2}\int\int \frac{e^{ik_b|\vec{r}-\vec{r}^{\,\prime}|}}{|\vec{r}-\vec{r}^{\,\prime}|}\varphi_{b}^{\ast}(\xi_{1}^{\prime}\ldots\xi_{N}^{\prime}) \nonumber \\
\times \sum_{t}t^{(i)}(\vec{r}^{\,\prime},\xi_{1}^{\prime}\ldots\xi_{N}^{\prime}) \mathscr{F}_{a}^{(i)} (\vec{r}^{\,\prime},\xi_{1}^{\prime}\ldots\xi_{N}^{\prime})d^{3}r^{\prime}d^{3}\xi_{1}^{\prime}\ldots d^{3}\xi_{N}^{\prime}\,.
\end{eqnarray}
Transformation of (\ref{4.40}) into a differential equation gives
\begin{eqnarray}
\label{4.41}
(\Delta_r+k_b^2)\psi_{ba}(\vec{r})-\frac{2m\gamma}{\hbar^2}\int\int \varphi_{b}^{\ast}(\xi_{1}^{\prime}\ldots\xi_{N}^{\prime}) \nonumber \\
\times\sum^{N}_{i=1}t^{(i)}(\vec{r},\xi_{1}\ldots\xi_{N}) F_{a}^{(i)} (\vec{r},\xi_{1}\ldots\xi_{N})d^{3}\xi_{1}\ldots d^{3}\xi_{N}=0
\end{eqnarray}
or
\begin{equation}
\label{4.42}
(\Delta_r+k_b^2)\psi_{ba}(\vec{r})-\frac{2m\gamma}{\hbar^2}\sum_f\sum^{N}_{i=1}t^{(i)}_{bf}(\vec{r}) F_{fa}^{(i)}(\vec{r})=0\,,
\end{equation}
where
\begin{eqnarray*}
    t^{(i)}_{bf}(\vec{r})=\int \varphi_{b}^{\ast}(\xi_{1}\ldots\xi_{N}) t^{(i)}(\vec{r},\xi_{1}\ldots\xi_{N})
    \varphi_{a}(\xi_{1}\ldots\xi_{N})d^{3}\xi_{1}\ldots d^{3}\xi_{N}\,; \\
    \mathscr{F}_{fa}^{(i)}(\vec{r})=\int
    \varphi_{f}^{\ast}(\xi_{1}\ldots\xi_{N})\mathscr{F}_{a}^{(i)}(\vec{r},\xi_{1}\ldots\xi_{N})d^{3}\xi_{1}\ldots d^{3}\xi_{N}\,.
\end{eqnarray*}
Let us consider in more detail the equation describing the elastically scattered wave:

\begin{eqnarray}
\label{4.43}
(\Delta_r+k_a^2)\psi_{aa}(\vec{r})-\frac{2m\gamma}{\hbar^2}\sum^{N}_{i=1}t^{(i)}_{aa}(\vec{r}) \mathscr{F}_{aa}^{(i)}(\vec{r}) \nonumber \\
-\frac{2m\gamma}{\hbar^2}\sum_{f\neq a}\sum^{N}_{i=1}t^{(i)}_{af}(\vec{r}) \mathscr{F}_{fa}^{(i)}(\vec{r})=0\,.
\end{eqnarray}
The amplitude $\mathscr{F}_{fa}^{(i)}$  in the third term (unlike
$\mathscr{F}_{aa}^{(i)}$) appears only as a result of rescattering
(see the general expression (\ref{4.38})). For this reason, under
the conditions when the elastic scattering amplitude $f$ is of the
same order of magnitude as the inelastic scattering amplitude and
much smaller than the distance between the scatterers, the third
term in the relation $f/R$ for correlated scatterers and
$(f/R)^{2}$ for independent scatterers is smaller than the second
term, and can be discarded.

As a result we have:
\begin{equation}
\label{1.98}
\mathscr{F}_{aa}^{(i)}(\vec{r})=\Phi_a (\vec{r})+ \frac{2 m \gamma}{\hbar^2} \sum _{l\neq i} \int G_{aa} (\vec{r}-\vec{\rho}')t_{aa}^{(l)} (\vec{\rho}')\mathscr{F}_{aa}^{(l)}(\vec{r}')d^3 (r')
\end{equation}
\begin{equation}
\label{1.99}
G_{aa}(\vec{r}-\vec{r}')= -\frac{1}{4 \pi} \frac{e^{ik_a |\vec{r}-\vec{r}'|}}{|\vec{r}-\vec{r}'|}
\end{equation}
%
The system of equations (\ref{4.43},\ref{1.98}) describes
propagation of the coherent elastic scattered wave in crystal.
Study of mentioned equations for the case of interaction between
thermal neutrons and photons with crystals was conducted in
\cite{b12,b30}. Specifically, it was shown that the effective
potential energy of interaction between thermal neutrons, which
possess de Broglie wavelength with the radius larger than the one
of nucleus ($S$-scattering), and the amplitude of thermal
oscillations of nucleus, is determined by more complex magnitude
than scattering amplitude:
\begin{equation}
\label{eq100}
F=\frac{f}{1+ikf}
\end{equation}
%
%
The stated result for crystal,neglecting thermal oscillations of
the nucleus, was obtained in \cite{b31}. The case of photons was
described in \cite{b12, b30}. Magnitude $F$ is the element of the
reaction matrix $K$ (for example see \cite{bn13,bn15}). Equations
(\ref{4.42}-\ref{4.43}), derived in \cite{b12, b32}, allow one to
consider the case of fast particles with de Broglie wavelength
comparable to or much smaller than atoms size and scattering
amplitude of nuclei in crystals. It is important to notice, that
for obtaining (\ref{eq8}-\ref{eq12}), the equation for
$\mathscr{F}_{aa}^{(i)}$ can be expressed as:

\begin{equation}
\label{eq101}
\mathscr{F}_{aa}^{(i)}(\vec{r})=\Psi_{aa} (\vec{r}) - \frac{2 m \gamma }{\hbar^2} \int G_{aa} (\vec{r}-\vec{r} ') t_{aa}^(i)(\vec{r} ')\mathscr{F}_{aa}^{(i)}(\vec{r} ')d^3 \vec{r} '
\end{equation}
%
As a result the equation for the coherent wave propagating through the crystal takes the form:
\begin{eqnarray}
\label{eq102}
&&(\Delta_r + k_a^2)\Psi_{aa}(\vec{r})-\frac{2 m \gamma}{\hbar^2} \sum_{i=1}^{N}t_{aa}^{(i)}(\vec{r})\Psi_{aa}(\vec{r})+ \nonumber \\
&&+\frac{2 m \gamma}{\hbar^2} \sum_{i=1}^{N}t_{aa}^{(i)}(\vec{r}) \frac{2 m \gamma}{\hbar^2} \cdot \int G_{aa} (\vec{r}-\vec{r} ') t_{aa}^{(i)}(\vec{r} ')\mathscr{F}_{aa}^{(i)}(\vec{r} ')d^3 r '=0.
\end{eqnarray}
%
Matrix elements of the scattering operator $t^{(i)}$ in the
equation (\ref{eq102}) act as effective interaction. This
interaction allows one to use the perturbation theory while
deriving (\ref{eq8}-\ref{eq12}) for both Coulomb, magnetic and
even strong interaction with nuclei in consequence of averaging
over nuclei oscillations in crystal. As a result, in the last term
of (\ref{eq102}) function  $\mathscr{F}_{aa}^{(i)}(\vec{r} ')$ can
be replaced with $\Psi_{aa} (\vec{r} ')$ so that the closed
equality for function $\Psi_{aa} (\vec{r})$ can be obtained.
\begin{eqnarray}
\label{eq103}
(\Delta_r + k_a^2)\Psi_{aa}(\vec{r})-\frac{2 m \gamma}{\hbar^2} \sum_{i=1}^{N} t_{aa}^{(i)} (\vec{r}) \Psi_{aa}(\vec{r})+ \nonumber \\
+ \frac{2 m \gamma}{\hbar^2} \sum_{i=1}^{N} t_{aa}^{(i)} (\vec{r}) \frac{2 m \gamma}{\hbar^2} \int G_{aa} (\vec{r}-\vec{r} ')\cdot\nonumber \\
\cdot t_{aa}^{(i)}(\vec{r}) \Psi_{aa}(\vec{r} ') d^3 r '=0.
\end{eqnarray}
%
Operators  $t^{(i)}_{aa}$ in equation (\ref{eq103}) depend on the coordinates of position of the center of the atom  $R_i$ :
$t^{(i)}_{aa}(\vec{r})=t^{(i)}_{aa}(\vec{r}-\vec{R}_i) $. Sums present in the equation (\ref{eq103}), are periodical functions of $\vec{r}$.
Let us introduce Fourier expansion for operator $t_{aa}^{(i)}$:
\begin{equation}
\label{eq104}
t_{aa}^{(i)} (\vec{r})=\frac{1}{(2 \pi)^3} \int t_{aa} (\vec{q}) e^{i \vec{q} (\vec{r}-\vec{R}_i)} d^3 q
\end{equation}
where $t_{aa}^{(i)} (\vec{q})$ does not contain index $i$ when lattice consists of atoms of one kind.

Summation over $\vec{R}_i$ in  (\ref{eq103}) and representation of
the Green function (\ref{eq21}) leads to the expression for
effective potential energy of interaction (\ref{eq8}-\ref{eq12}).

 \end{document}